\documentclass{aip-cp}

\usepackage[numbers]{natbib}
\usepackage{rotating}
\usepackage{graphicx}
\usepackage{tabularx}
\begin{document}

\title{Scalar Field Condensate
Baryogenesis Model in Different Inflationary Scenarios}

\author[aff1]{Daniela Kirilova\corref{cor1}}
\author[aff1]{Mariana Panayotova}
\eaddress{mariana@astro.bas.bg}

\affil[aff1]{Institute of Astronomy with NAO Rozhen, Bulgarian
Academy of Sciences, 72 Tsarigradsko shose Blvd., Sofia, Bulgaria} 
\corresp[cor1]{Corresponding author: dani@astro.bas.bg}

\maketitle

\begin{abstract}
We calculate the baryon asymmetry value generated in the Scalar
Field Condensate (SCF) baryogenesis model obtained in several
inflationary scenarios and different reheating models. We provide
analysis of the baryon asymmetry value obtained for more than 70
sets of parameters of the SCF model and the following inflationary
scenarios, namely: new inflation, chaotic inflation, Starobinsky
inflation, MSSM inflation, quintessential inflation. We considered
both cases of efficient thermalization after inflation and also
delayed thermalization.

We have found that SFC baryogenesis model produces baryon asymmetry
by orders of magnitude bigger than the observed one for the
following inflationary models: new inflation, new inflation model by
Shafi and Vilenkin, MSSM inflation, chaotic inflation  with high
reheating temperature and the simplest Shafi-Vilenkin chaotic
inflationary model. For these models strong diluting mechanisms are
needed to reduce the resultant baryon excess at low energies to its
observational value today.

We have found that a successful generation of the observed baryon
asymmetry is possible by SCF baryogenesis model in Modified
Starobinsky inflation, chaotic inflation with low reheating
temperature, chaotic inflation in SUGRA and Quintessential
inflation.

\end{abstract}

\section{1.Introduction}

The inflationary paradigm for the description of the very early
Universe is already more than 30 years old, however, there still
exist numerous models of inflation ~\cite{Linde04}. Moreover, the
reheating process at the end of inflation, that is believed to have
provided the transfer of the energy stored in the inflaton to other
fields and thus enabled the beginning of the radiation dominated
stage of the Universe, could have also proceeded through different
mechanisms (perturbative \cite{DK90}, nonperturbative \cite{FKL99}
(see also the discussion of different reheating mechanisms in
ref.~\cite{PhT})). Also different decay channels and different decay
rates of the inflaton and other particles, and different
thermalization (instant or delayed) are possible \cite{MZ14}.
Therefore, it is interesting to consider the possibility for
production of the observed baryon asymmetry $\beta=6\times10^{-10}$
for different reheating temperatures in different inflationary
models.

In this work we analyze the baryon asymmetry generation according to
the Scalar Field Condensate (SCF) baryogenesis model for several
inflationary scenarios and different reheating models.

\subsection{1.1 Baryon Asymmetry of the Universe}

The generation of the baryon asymmetry of the Universe is one of the
open cosmological issues.  Both cosmic ray data and gamma ray data
indicate that there are no significant quantities of antimatter in
the local vicinity up to galaxy cluster scales of 10-20
Mpc~\cite{Steigman76, Steigman08, Stecker85, Ballmoos14, Dolgov15}.
It is most probable that our Universe is made of matter.

 The
baryon asymmetry is usually described by:
\begin{equation}
\beta=(N_b-N_{\bar{b}})/N_{\gamma}\sim N_b/N_{\gamma}=\eta,
\end{equation}
where $N_b$ is the number of baryons, $N_{\bar{b}}$ is the number of
anti-baryons, $N_{\gamma}$ - the number of photons.

The baryon-to-photon ratio $\eta$ is precisely measured today,
namely:
\begin{equation}
\eta \sim 6\times 10^{-10},
\end{equation}

\noindent the best baryometers being BBN and CMB measurements.
 Among the light elements produced during BBN Deuterium is the most
 sensitive to $\eta$, thus the most precisely obtained $\eta$,
 based on BBN theory and D observations is~\cite{PC12}:

 $$\eta_D=6 \pm 0.3 \times 10^{-10} ~~\rm{at}~~  95\% ~C.L.$$

 The CMB anisotropy data measures $\eta$ with comparative accuracy,
 namely (see ref.\cite{Planck16}):
 $$\eta_{CMB}=6.11 \pm 0.04 \times 10^{-10}~~ \rm{at}~~  68\%~ C.L.$$

 Though these independent measurements correspond to quite different
 epochs, namely BBN proceeds at $z \sim 10^9$, while CMB to $z \sim 1000$ they
 are in excellent agreement, t.e. there was no change in this ratio
 between the two epochs.

 At present there exist many baryogenesis models, which
 successfully generate this number at quite different epochs, in the
 wide range between the end of inflation and before BBN.
 Just to mention the most popular ones: GUT baryogenesis~\cite{Sakharov67, KRSh85}, SUSSY
 baryogenesis, baryogenesis through leptogenesis~\cite{FY86}, Afleck and Dine
 baryogenesis~\cite{AD85}, Scalar Field Condensate baryogenesis (SFC)~\cite{DK90, DK91}, etc.

 In the next section we provide a short description of the SFC
 baryogenesis model. In the third section we present our results for the calculated
baryon asymmetry $\beta$ in different inflationary scenarios and
different reheating models. In the conclusion we list the main
results and present a short discussion.

\subsection{2. SFC short description}

First ideas on SFC baryogenesis model and analytical construction of
that model were presented in refs.~\cite{DK90, DK91}.
In following publications an inhomogeneous SFC baryogenesis model
was explored semi-analytically, and was applied to explain the very
large scale structure in the universe and the quasi-periodicity
found at very large scales with typical period of 128 $h^{-2}$
Mpc~\cite{CK96, CK00, K03}.
Since first analytical considerations~\cite{DK90} it is known that
particle creation processes play important role for the
determination of the baryon asymmetry generation in that model.
Recently, more precise numerical account for particle creation
processes and their role in SFC baryogenesis was provided in refs.
~\cite{KP14, KP15, PK16}. According to SCF baryogenesis model at the
inflationary stage there existed the inflaton $\psi$ and a complex
scalar field $\varphi$, carrying baryon charge. During inflation, as
a result of the rise of quantum fluctuations of $\varphi$, a
condensate $<\varphi>\neq0$ with a nonzero baryon charge $B$ is
formed ~\cite{VF82, BD78, S82}. $B$ is not conserved at large
$\varphi$ due to the presence of B non-conserving (BV)
self-interaction terms in the potential V($\varphi$).

The equation of motion of $\varphi$ is:
\begin{equation} \ddot{\varphi}+3H\dot{\varphi}+
{1 \over4} \Gamma_{\varphi}\dot{\varphi}+ U'_{\varphi}=0,
\end{equation}
where $a(t)$ is the scale factor, $H$ is the Hubble parameter
$H=\dot{a}/a$. $\Gamma_{\varphi}=\alpha\Omega$   is the rate of
particle creation, $\Omega = 2\pi/T$, where T is the period of the
field oscillations.  The analytically estimated value: $\Omega_0 =
\lambda^{1/2} \varphi_0$, is used as an initial condition of the
frequency.

\begin{equation} B = - i ({{\varphi}'}^{\ast}{\varphi}-{\varphi}'{\varphi}^{\ast})
\end{equation}

The potential is chosen of the form:
\begin{equation} U(\varphi)=m^2\varphi^2+{\lambda_1\over 2}|\varphi|^4+
{\lambda_2\over 4}(\varphi^4+\varphi^{*4})+ {\lambda_3\over
4}|\varphi|^2(\varphi^2+\varphi^{*2}). \end{equation}

The following natural assumptions are made: the mass parameters of
the potential are small in comparison with the Hubble parameter
during inflation $m \ll H_I$, the self-coupling constants
$\lambda_i$ are of the order of the gauge coupling constant $\alpha$
and $m$ is in the range $10^2 - 10^4$ GeV. The energy density of
$\varphi$ at the inflationary stage is of the order $H^4_I$, hence
\begin{equation} \varphi^{max}_o \sim H_I\lambda^{-1/4} ,  \quad
\dot{\varphi_o}=(H_I)^2 ,  \quad B_0=H_I^3.
\end{equation}

At the end of inflationary stage $\varphi$ starts to oscillate
around its equilibrium and its amplitude decreases due to the
Universe expansion and the particle creation processes, resultant
from the coupling of the scalar field to fermions  $g\varphi
f_1f_2$, where $g^2/4\pi = \alpha_{ GUT}$. In SFC baryogenesis model
$B$, contained in the condensate, can be considerably reduced due to
particle creation at the BV stage \cite{DK91, KP07}. Therefore,  at
the high energy stage, where baryon violation is considerable, $B$,
contained in $\varphi$ condensate, is reduced due to particle
production.

 Here we provide
numerical account for the particle creation processes of $\varphi$.

BV becomes negligible at small $\varphi$. $B$ which survives until
B-conservation epoch $t_B$, is transferred to fermions and the
excess of matter is produced.

We have provided numerical analysis \cite{KP07, KP12, KP14, KP15} of
the evolution of $\varphi(t)=x+iy$ and $B(t)$ from the inflationary
stage until $t_B$. We
 developed a computer program in Fortran 77 using Runge-Kutta 4th order method.
The system of ordinary differential equations, corresponding to the
equation of motion for the real and imaginary part of $\varphi$ and
B contained in it was solved calculating $\Omega$ at each step. The
numerical analysis included around 100 sets of parameters in their
natural ranges of values: $\alpha = 10^{-3} - 5 \times 10^{-2}$,
$H_I = 10^7 - 10^{12}$ GeV, $m = 100 - 1000$ GeV, $\lambda_1 =
10^{-3} - 5 \times 10^{-2}$, $\lambda_{2,3} = 10^{-4} - 5 \times
10^{-2}$. All considered in our calculations $H_I$ values are in
agreement with the observational constraint from Planck data,
namely: $H_I<3.7 10^{-5}M_{Pl}/(8\pi)^{1/2}$.

For each set of SFC baryogenesis model parameters we have calculated
the final B contained in the condensate $\varphi(t)$ before its decay.
The dependence of the produced B on the parameters of the models
(namely $m$, $H_I$, $\lambda_i$ and $\alpha$) were revealed.

The produced baryon asymmetry $\beta=(N_b-N_{\bar{b}})/N_{\gamma}$
 in SFC baryogenesis model depends on the generated baryon excess $B$, the
 reheating temperature of the Universe $T_R$ and the value of the Hubble parameter
 at the end of inflation $H_I$. Namely:
\begin{equation}
 \beta \sim N_B/T_R^3 \sim B T_R/H_I.
\end{equation}

  $T_R$ and $H_I$ values depend on the kind of inflation and
reheating.

Hence, in the present work we calculated the baryon asymmetry of the
Universe produced in the SFC baryogenesis model using the available
results on B for all studied range of model's parameters from
ref.~\cite{KP15} and considered different
 models of inflation and reheating.
 In the next section we present the results of our
analysis for the generated baryon asymmetry $\beta$ in several
inflationary models and different reheating scenarios.

\section{3.Baryon Asymmetry in Different Inflationary Models}

\subsection{3.1.Notes on Inflation and Reheating}

 The idea of an exponential inflationary stage in the early
evolution of the Universe has been established  as an extension to
the standard cosmological model in order to resolve several
conceptional problems of the standard cosmological model, among
which homogeneity, isotropy, flatness of the Universe and the
predicted over abundance of magnetic
monopoles. 

Now there exist hundreds models of inflation (see for example the
Encyclopaedia Inflationaris collection \cite{Mar}). 
Recently the inflation models were probed by the Planck data. Planck
2013, 2015, and 2018 releases have put strong constraints  on
several types of inflationary  models.

Chronologically, the first realistic inflationary model was created
by Starobinsky in 1980~\cite{S80}. The Starobinsky $R^2$ inflation
model has a potential as follows:
\begin{equation}
V(\psi) = \Lambda^4 \left( 1 - e^{- \sqrt{2/3} \psi/M_\mathrm{pl}
}\right)^2 \label{R2_Einsteinframe}
\end{equation}

 The model is in a good agreement with Planck18 data.

In 1981  Linde  \cite{L82} and Albrecht and Steinhardt \cite{AS82}
independently proposed a {\it new inflation} or slow-roll inflation
model, where  inflation occurred by a scalar field rolling down a
potential energy hill, instead of tunneling out of a false vacuum
state, as in ref. \cite{Guth81}.


In 1983 the {\it chaotic inflationary model} was proposed, which
does not require an initial state of thermal equilibrium,
supercooling and tunneling from the false vacuum. This class of
inflationary models has a single monomial potential  \cite{L85}:
\begin{equation}
V(\psi) = \lambda M_\mathrm{pl}^4 \left( \frac{\psi}{M_\mathrm{pl}}
\right)^p \,, \label{PowerLawPot:Eq}
\end{equation}
where inflation occurs at $\psi > M_\mathrm{pl}$. Planck18 data
disfavors potentials with $p \geq 2$ but models with simple linear
potentials $p=1$ or $p=2/3$ and fractional power monomials are more
compatible.

Other popular inflationary model is the model of {\it quintessential
inflation} of Peebles and Vilenkin \cite{PV99}, which provides a
unified description for both the inflationary stage and the current
acceleration stage of the Universe using a single scalar field
potential:
\begin{equation}
V=\lambda (\psi^4 + M^4), \quad\psi < 0,
\end{equation}
\begin{equation}
V={\lambda M^8\over{\psi^4 + M^4}},\quad\psi\geq 0. \label{5}
\end{equation}

At $-\psi\gg M$ this is a “chaotic“ inflation potential \cite{L85},
at $\psi\gg M$ it is a “quintessence“ form, $\lambda = 1\times
10^{-14}$. Some model improvements were proposed lately to obtain
agreement with the recent Planck18 observational data \cite{HAP19}.


 Planck CMB anisotropy measurements~\cite{Planck}
 put constraints on inflationary models.

We have considered here the  following inflationary models: the new inflation \cite{L82, AS82}, Shafi-Vilenkin model of new
inflation, chaotic inflation \cite{L85, L90}, Shafi-Vilenkin model
of chaotic inflation, chaotic inflation in SUGRA, Starobinsky
inflation \cite{KLS85}, MSSM inflation and quintessential inflation.

 Besides the great variety
of inflationary models, there exist also different possibilities for
reheating realizations after inflation.
 During reheating the inflaton energy is transferred to other
 dynamical degrees of freedom, which results in radiation dominated
 stage of the Universe. However, there exist different reheating
 mechanisms~\cite{PhD}. The resultant $T_R$ depends on the way reheating proceeds:
 namely reheating
by perturbative decay of the inflaton $\psi$ and by non-perturbative
decay of the inflaton $\psi$, it depends also on the inflaton decay
rate, on the spectrum of inflaton decay particles, on the
thermalization after inflation (instantaneous or
 delayed) reheating, etc.~\cite{marko}.

 There exist CMB and BBN constraints on the inflationary reheating temperature.
 Reheating should proceed before BBN and  $T_R>5$  MeV, so that low
 reheating temperature would not effect strongly the properties of
 neutrino and consequently BBN He production and CMB~\cite{martin10, salas}.

 On the other hand, reheating should proceed at low enough energy so
 that GUT symmetry is not restored and thus monopole problem is
 evaded. Besides in SUSY models in order to avoid gravitino
 overproduction, which can destroy BBN predictions the reheating
 temperature should be $T_R<10^7-10^9$ GeV. This is a constraint in case gravitinos are in the mass range (100 GeV - 1 TeV)~\cite{TR}.
 When gravitino mass is $>10$ TeV, another constraint holds:
 $T_R<10^{11}$ GeV, following from the constraints on gravitino
 number from overclosure bound~\cite{gravitino}.


 In the 90ies of the previous century the preheating by
perturbative decay of the inflaton $\psi$ into fermions was
considered. See the pioneer works of refs.~\cite{DK90,TB90}.
Reheating takes place when $H$ drops to the value of $\Gamma$, the
total decay rate of the inflaton, and inflaton decay becomes
effective. Using the Friedmann equation and assuming an {\it
instantaneous} conversion of the inflaton energy at the end of
inflation into radiation $\rho_r=g_* \pi^2/30 T^4=\rho_{\psi}$ and
{\it fast thermalization}, the reheating temperature is given by:
\begin{equation}
T_R=(90/8\pi^3 g_*)^{1/4}(M_{Pl}H)^{1/2}.
\end{equation}
Reheating completes when $H$ becomes less than $\Gamma/2$. Then an
upper bound is obtained for $\Gamma=2 H$ in case of efficient
thermalization, namely:
\begin{equation}
T_R=(90/32\pi^3 g_*)^{1/4}(M_{Pl}\Gamma)^{1/2},
\end{equation}
where  $g_*$ is of the order $10^2$, $T_R\sim
0.1(M_{Pl}\Gamma)^{1/2}$. Then the typical $T_R$ is less than $10^9$
GeV~\cite{KLS94, KLS97}.

However, $\psi$ may decay into other bosons due to broad resonance
~\cite{KLS94, KLS97, Boy95}. In this case $T_R$ may be $\sim
10^{12}$ GeV, i.e. much higher than $T_R$ estimated in eq.13.
Non-perturbative preheating was discussed for example in
refs.~\cite{Al11, FKL98, F95, ABM00}.

On the other hand, $T_R$ may be much smaller than these estimations
in case of {\it slow thermalization} when local thermodynamical
equilibrium is not reached until the beginning of the RD epoch. This
is usually the case for small inflaton couplings and very big
inflaton masses. The conditions for efficient or inefficient
thermalization were discussed in ref.~\cite{MZ14}.

Thus, there is a large range of the allowed values of $T_R$. We have
used in our analysis $T_R$ in the range $[10^5, 10^{14}]$ GeV.

\section{3.2 Baryon Asymmetry in Different Inflationary Models - Results}

  Here we
present our results of the baryon asymmetry value calculated for
different reheating possibilities and different inflationary
scenarios. First consideration of SFC baryogenesis model in
different inflationary scenarios and preliminary results were
reported in ref.~\cite{AIP, BS}. In the present work we considered all
$B$-excess values in the whole range of studied parameter sets of
the SFC baryogenesis model.

The results of our analysis on the calculated $\beta$ for certain
$T_R$ values corresponding to different inflationary scenarios and
different types of thermalization are presented below:

\subsection{3.2.1 Inflationary models with overproduction of baryon asymmetry}

 In case of {\it new inflation} \cite{DL82, AS82} for
$H_I=10^{10} $ GeV and $T_R=10^{14}$ GeV we have found that the
calculated baryon asymmetry for all sets of model's parameters is
several orders of magnitudes bigger than the observational value
$\beta_{obs}$. (The same holds if one varies $H_I$ within the range
$[5\times 10^9, 5
\times 10^{10}]$ GeV.)

In the new inflation model by Shafi and Vilenkin \cite{SV84} for
$H_I=3 \times 10^9$ GeV and $T_R=3\times10^7$ GeV, the calculated
baryon asymmetry again is much bigger than $\beta_{obs}$, namely
$\beta>10^{-7}$. (We have calculated $\beta$ for all sets of the SFC
model parameters and for $H_I$ in the range $[10^8-10^{10}]$ GeV.)

In case of {\it chaotic inflation}, for $H_I \in [10^{11},10^{12}]$
GeV, and $T_R<3\times10^{14}$ GeV, namely we have used $T_R \in
[10^{12},10^{14}]$ GeV , the calculated asymmetry is $\beta
>10^{-5}$.

For the simplest Shafi-Vilenkin model in chaotic inflation
$T_R=10^{12}-10^{13}$ GeV again $\beta >10^{-7}$. We have provided
the analysis for $H_I \in [5 \times 10^9, 10^{12}]$ GeV.

However, depending on the inflaton decay rate, decay channels,
couplings, $T_R$ may be lower, namely $T_R \in [10^8,10^{11}]$ GeV.
$T_R$ may be lower, thus allowing successful baryogenesis in these
chaotic inflationary models, i.e. even in case of big $H_I\sim
10^{12}$ GeV $\beta_{obs}$ can be obtained (as will be discussed in
the next subsection).

 In MSSM inflation model \cite{A06, F17} with $H_I=1$ GeV, $T_R=2\times 10^8$ GeV,  SCF baryogenesis model
does not work, $\beta >> \beta_{obs}$. This inflationary model also
has severe problems with gravitino overproduction, violating BBN and
observed DM abundance.

\subsection{3.2.1 Inflationary models with successful production of the observed baryon asymmetry}

We have used the values of $T_R$ and $m_{\psi}$ for
$\alpha_{\psi}=10^{-11}$ from ref.~\cite{MZ14} to calculate the
baryon asymmetry for the different cases, different $T_R$.( Assuming
$\Gamma=\alpha_{\psi}m_{\psi}$ the authors have calculated
$T_R(m_{\psi})$  from ref.~\cite{AFW82} for different
$\alpha_{\psi}$ and for delayed and efficient thermalization.)

In case of {\it efficient thermalization} we have found that the
production of the observed value of $\beta$ is possible for
$H_I=10^{12}$ GeV, $T_R=6.2 \times 10^9$ GeV and specific sets of
SFC model's parameters, namely: $\lambda_1=5 \times 10^{-2}$,
$\alpha \in [3 \times 10^{-2},5 \times10^{-2}]$, $\lambda_2=\lambda_3
\in [10^{-3},10^{-2}]$, $m=350$ GeV. For $H_I=10^{11}$ GeV and
$T_R=1.9 \times 10^9$ GeV the production of the observed value of
$\beta$ is possible when $\lambda_1=\alpha=5 \times 10^{-2}$,
$\lambda_2=\lambda_3=10^{-2}$ and $m=100, 200$ and $350$ GeV.

In case of {\it delayed thermalization} there appear possibilities
for the production of $\beta_{obs}$, corresponding to $H_I=10^{12}$
GeV, $T_R=4.5 \times 10^8$ GeV and several different sets of model's
parameters in the following ranges $\lambda_1 \in [5 \times 10^{-2},
10^{-2}]$; $\alpha \in [5 \times 10^{-2},10^{-2}]$,
$\lambda_2=\lambda_3=10^{-3}$ and $m=350-500$ GeV. It could be
easily seen that the strongest influence comes from $\alpha$
parameter of the model and then from $m$. For fine tuning
$\lambda_1$ can be used and then $\lambda_2= \lambda_3$ parameters
of the SCF model. (We have studied $H_I$ in the range
$[10^7-10^{12}]$ GeV.)

In case of modified {\it Starobinsky inflation}  \cite{S80} $T_R=0.1
(\Gamma M_{Pl})^{1/2}=10^9$ GeV, $H_I=10^{11}$ GeV, successful
baryogenesis is possible for the efficient thermalization as well.
Namely $\beta =\beta_{obs}$ was found possible for several sets of
model's parameters.  (We have studied $H_I$ in the range $[5 \times
10^6, 10^{12}]$ GeV.) For the simplest extension of the Starobinsky
inflation see ref.~\cite{Bruck16}.

For monomial potential of eqn. 8
we have calculated $\beta$ for $p=2/3$ and $T_R\in [10^8,10^{11}]$
GeV. For $T_R=10^9$ GeV and $H_I \sim 10^{11}$ GeV $\beta_{obs}$ can
be produced.

For {\it chaotic inflation in SUGRA} \cite{NOS83} $T_R>10^9$ GeV  it
is possible to generate $\beta_{obs}$.

In {\it quintessential inflation} with $T_R=2 \times 10^{5}$ GeV and
decay into massless particles, the production of the baryon
asymmetry value is successful for $H_I=10^{12}$ GeV and several sets
of SCF model parameters as well. When  $m=350$ GeV, the favored
value of $\alpha$ parameter is $\alpha=10^{-3}$ with wider range of
$\lambda_1 \in [10^{-3},5 \times 10^{-2}] $ and $\lambda_2=\lambda_3
\in [10^{-4}, 5 \times 10^{-3}]$. Again, strong dependance of
$\alpha$ parameter may be mentioned since $\alpha$ reflects on the
time of the scalar field decay and therefore it has influence on the
baryon excess and baryon asymmetry values.

The particular parameters sets of the SCF baryogenesis model and the
inflationary models and the types of thermalization, for which
successful production of the baryon asymmetry close to its
observational value is possible, are listed in the table.

\begin{table}[h!]
\centering
\begin{tabularx}{1\textwidth} {
  | >{\raggedright\arraybackslash}X
  | >{\raggedright\arraybackslash}X
  | >{\raggedright\arraybackslash}X
  | >{\raggedright\arraybackslash}X | }

\hline
Starobinsky Inflation
& $H_I=10^{11}$ GeV; $T_R=10^9$ GeV
& $\lambda_1=\alpha=5\times10^{-2}$, $\lambda_2=\lambda_3=10^{-2}$, $m=100$ GeV, $\beta=9.3\times10^{-10}$
&\\
\hline
& $H_I=10^{12}$ GeV; $T_R=10^9$ GeV
& $\lambda_1=5\times10^{-2}$, $\alpha=3\times10^{-2}$, $\lambda_2=\lambda_3=10^{-3}$, $m=350$ GeV, $\beta=6.6\times10^{-10}$
& $\lambda_1=\alpha=5\times10^{-2}$, $\lambda_2=\lambda_3=10^{-3}$, $m=350$ GeV, $\beta=8.0\times10^{-10}$ \\
\hline
\hline

Quintessential Inflation
& $H_I=10^{12}$ GeV; $T_R=2 \times 10^5$ GeV
& $\lambda_1=5\times10^{-3}$, $\alpha=10^{-3}$, $\lambda_2=\lambda_3=10^{-4}$, $m=350$ GeV, $\beta=4.6\times10^{-10}$
& $\lambda_1=10^{-2}$, $\alpha=10^{-3}$, $\lambda_2=\lambda_3=10^{-4}$, $m=350$ GeV, $\beta=7.8\times10^{-10}$ \\
\hline
\hline

Chaotic Inflation, Efficient Thermalization
& $H_I=10^{12}$ GeV;
$T_R=6.2 \times 10^9$ GeV & $\lambda_1=\alpha=5\times10^{-2}$, $\lambda_2=\lambda_3=10^{-2}$, $m=350$ GeV, $\beta=7.4\times10^{-10}$
& \\
\hline
\hline

Chaotic Inflation, Delayed Thermalization
& $H_I=10^{12}$ GeV; $T_R=4.5 \times 10^8$ GeV
& $\lambda_1=\alpha=10^{-2}$, $\lambda_2=\lambda_3=10^{-3}$, $m=350$ GeV, $\beta=9.5\times10^{-10}$
& $\lambda_1=\alpha=5\times10^{-2}$, $\lambda_2=\lambda_3=10^{-3}$, $m=350$ GeV, $\beta=3.6\times10^{-10}$ \\
\hline

\end{tabularx}
\caption{Successful production of the observed baryon asymmetry
value $\beta_{obs}$ for particular sets of SCF model parameters in
different inflationary scenarios} \label{table1}
\end{table}

As it can be seen from the Table \ref{table1} in case of Starobinsky and chaotic
inflationary scenarios with successful production of the baryon
asymmetry value for $T_R \in [4.5\times 10^8, 6.2 \times 10^9]$ GeV
and $H_I \in [10^{11}, 10^{12}]$, the SFC parameters can be
approximately fixed - they lie within the following ranges:  $m \sim
350$ GeV (with one exception) $\alpha \in [10^{-2}, 5 \times
10^{-2}]$, $\lambda_1 \sim 5 \times 10^{-2}$, $\lambda_{2,3} \in
[10^{-3}, 10^{-2}]$.

In case of quintessential inflation, however, for $H_I \sim 10^{12}$
GeV and $m \sim 350$ GeV, the reheating temperature is much lower
$T_R \sim 2 \times 10^5$ GeV and the rest of the SFC parameters -
the coupling constants have lower values, namely $\alpha \sim
10^{-3}$, $\lambda_1 \in [5 \times 10^{-3}, 10^{-2}]$,
$\lambda_{2,3} \in [10^{-4}, 5 \times 10^{-5}]$.

Quintessential inflationary models need the smallest $\lambda_i$ and
$\alpha$, by order of magnitude smaller than the ones in other
considered here inflationary models. If we impose the requirement
for the value of $\alpha$ to be close to the $\alpha_{GUT}$, SFC
baryogenesis cannot be realized in Quintessential inflation.

Vice versa, fixing the inflationary model we can fix the SFC model
parameters.

\begin{figure}[h]
  \centerline{\includegraphics[width=1. \textwidth ]{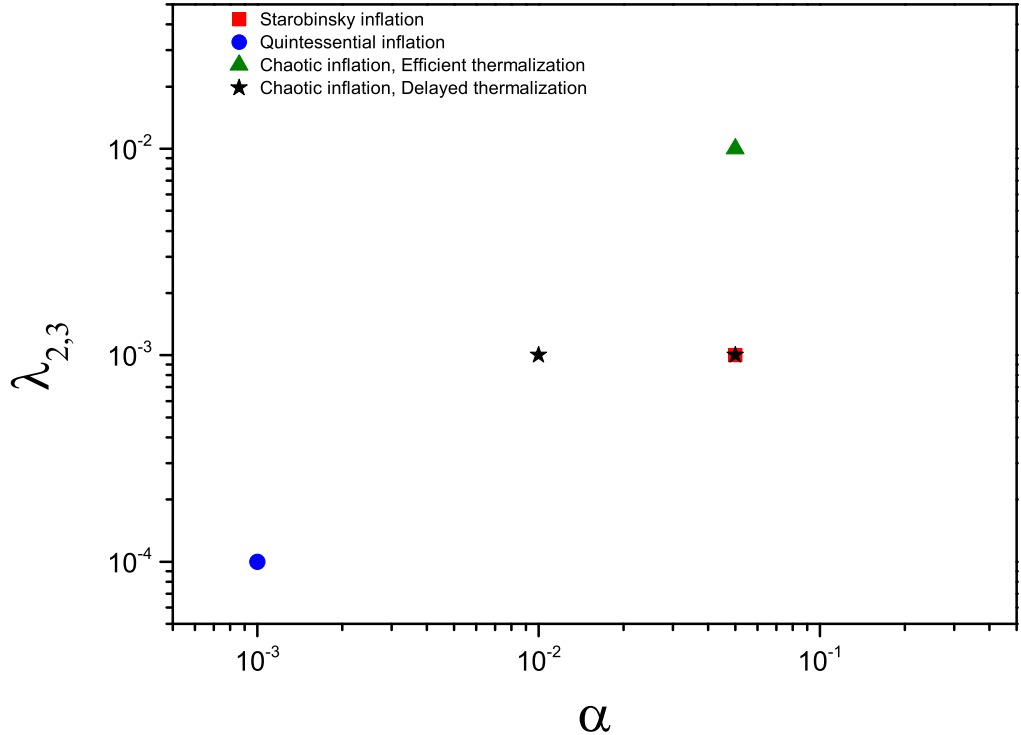}}
  \caption{ The figure presents different inflationary
models in the $\alpha$-$\lambda_{2,3}$ plane for which successful
SFC baryogenesis is achieved for the following parameters:
$\lambda_1=5 \times 10^{-2}, m=350$ GeV and $H_I=10^{12}$ GeV.}
\label{fig1}
\end{figure}

 In Figure \ref{fig1} we present in
the $\alpha$-$\lambda_{2,3}$ plane the inflationary models in which
the closest to the observational baryon asymmetry value is
generated. The other fixed values for model parameters are:
$\lambda_1=5 \times 10^{-2}, m=350$ GeV and $H_I=10^{12}$ GeV. Mind
that,  however, the models correspond to different reheating
temperatures, as given in Table \ref{table1}.

We have also results for SUGRA inflationary model. In case of $T_R
\sim 10^{12}$ GeV the obtained $\beta$ are bigger than the observational
value independent of the $H_I$ value assumed. For $T_R=10^9$ GeV, the
results coincide with the ones for Starobinsky model.

\section{Conclusions}

On the basis of the numerical analysis of the evolution of baryon
charge $B(t)$ produced in the SFC baryogenesis model and the
estimation of the produced baryon asymmetry for different sets of
models parameters and different reheating temperatures of several
inflationary scenarios we have shown that:

(i) SFC baryogenesis model produces baryon asymmetry by orders of
magnitude bigger than the observed one for the following
inflationary models: new inflation, new inflation model by Shafi and
Vilenkin,  chaotic inflation  with high reheating temperature, the
simplest Shafi-Vilenkin chaotic inflationary model and MSSM
inflation.

For these models SCF baryogenesis needs strong diluting mechanisms
in order to reduce the resultant baryon excess at low energies to
its observational value today.

(ii) SFC baryogenesis model produces similar to the observed value
of the baryon asymmetry in the following inflationary models:
Modified Starobinsky inflation, chaotic inflation with lower
reheating temperature, chaotic inflation in SUGRA and
 Quintessential
inflation. In case of delayed thermalization, when  $T_R$ is much
lower, a successful SFC baryogenesis may be achieved more easily in
the chaotic inflationary models. Curiously enough these are also
models preferred by the Planck CMB data analysis.

However, choosing for the value of $\alpha$ the value closest to the
$\alpha_{GUT}$, it is possible to conclude that SFC baryogenesis
cannot be realized in Quintessential inflation.

Vice versa, fixing the inflationary model we can fix the SFC model
parameters.

 Encouraged by the numerous possibilities of successful $\beta$ generation found for
 the discussed in this work
 inflationary models, we consider it interesting to continue
the study SFC baryogenesis model in other
 inflationary models and also to expand the numerical analysis
 towards higher values of $m$ and $H_I$.
 The latter is necessary because there exist models, like power law inflation, braneworld
 inflation, quintessential power law inflation, etc. 
  which require higher than the upper bound used in our analysis
 values of $H_I$, i.e. higher than $H_I \sim 10^{12}$ GeV.

\section{ACKNOWLEDGMENTS}
We acknowledge the partial financial support by project
DN18/13-12.12.2017 of the Bulgarian National Science Fund of the
Bulgarian Ministry of Education and Science.


\begin{thebibliography}{1}

\bibitem{Linde04}
Linde A., Phys. Scripta T117, 40, (2005).
\bibitem{DK90}
Dolgov A., Kirilova D., Sov. J. Nucl. Phys. 51, 172, (1990).
\bibitem{FKL99}
Felder G., Kofman L., Linde A., JHEP 0002, 027, (2000).
\bibitem{PhT}
Moghaddam H., PhD thesis "Reheating in the Early Universe Cosmology", (2017).
\bibitem{MZ14}
Mazumdar A., Zaldivar B., Nuclear Physics B886, 312, (2014).
\bibitem{Steigman76}
Steigman G., Ann. Rev. Astron. Astrophys., 14, 339, (1976).
\bibitem{Steigman08}
Steigman G., J. Cosmol. Astropart. Phys., 0910, 001, (2008).
\bibitem{Stecker85}
Stecker F., Nucl. Phys. B, 252, 25, (1985).
\bibitem{Ballmoos14}
Ballmoos P., Hyperfine Interact., 228, 91, (2014).
\bibitem{Dolgov15}
Dolgov, A., EPJ Web of Conferences, 95, 03007, (2015).
\bibitem{PC12}
 Pettini M., Cooke R., Mon. Not. Roy. Astron. Soc., 425, 2477, (2012).
\bibitem{Planck16}
Ade P., et.al. [Planck Collaboration], Astron. Astrophys., 594, A13, (2016).
\bibitem{Sakharov67}
Sakharov A., JETP, 5, 32, (1967).
\bibitem{KRSh85}
Kuzmin V., Rubakov V., Shaposhnikov M., Phys. Rev. Lett. 84, 3756, (1985).
\bibitem{FY86}
Fukugita M., Yanagida T., Phys. Lett. B174, 45, (1986).
\bibitem{AD85}
Affleck I., Dine M., Nucl. Phys. B, 249, 361, (1985).
\bibitem{DK91}
Dolgov A., Kirilova D., J. Moscow Phys. Soc. 1, 217, (1991).
\bibitem{CK96}
Chizhov M., Kirilova D., AATr, 10, 69, (1996).
\bibitem{CK00}
Kirilova D., Chizhov M., MNRAS, 314, 256, (2000).
\bibitem{K03}
Kirilova D., Nucl. Phys. Proc. Suppl. 122, 404, (2003).
\bibitem{KP14}
Kirilova D., Panayotova M., BAJ 20, 45, (2014).
\bibitem{KP15}
Kirilova D., Panayotova M., Advances in Astronomy, 465, (2015).
\bibitem{PK16}
Panayotova M., Kirilova D., Bulg. J. Phys. 43, 327–333, (2016).
\bibitem{VF82}
Vilenkin A., Ford L., Phys. Rev. D26, 1231, (1982).
\bibitem{BD78}
Bunch T., Davies P., Proc. R. Soc. London, Ser. A 360, 117, (1978).
\bibitem{S82}
Starobinsky A., Phys. Lett. B117, 175, (1982).
\bibitem{KP07}
Kirilova D., Panayotova M., Bulg. J. Phys. 34 s2, 330, (2007).
\bibitem{KP12}
Kirilova D., Panayotova M., Proc. 8th Serbian-Bulgarian Astronomical Conference (VIII SBGAC), Leskovac, Serbia 8-12 May, (2012).
\bibitem{Mar}
Martin J., Ringeval C., Vennin V., Phys. Dark Univ.5-6, 75, (2014);
Martin J., Ringeval C., Trotta R., Vennin V., JCAP 1403, 039, (2014).
\bibitem{S80}
Starobinsky A., Phys.Lett. B91, 99, (1980).
\bibitem{L82}
Linde A., Phys. Lett. B108, 389, (1982).
\bibitem{AS82}
Albrecht A., Steinhardt P., Phys. Rev. Lett. 48, 1220, (1982).
\bibitem{Guth81}
Guth A., Phys. Rev. D 23, 347, (1981).
\bibitem{L85}
Linde A., Phys. Lett., 129B, 177, (1983); Phys. Lett., 162B, 281, (1985).
\bibitem{PV99}
Peebles P., Vilenkin A., Phys.Rev. D59, 063505 (1999).
\bibitem{HAP19}
Haro J., Amoros J., Pan S., Eur. Phys. J. C79, 505  (2019).
\bibitem{Planck}
Akrami Y., et al. [Plank Collaboration], Astronomy\&Astroph. 641, A10, (2020).
\bibitem{L90}
Linde A., \emph{Particle Physics and Inflationary Cosmology}, Harwood, Chur, Switzerland, (1990).
\bibitem{KLS85}
Kofman L., Linde A., Starobinsky A., Phys. Lett. B157, 36, (1985).
\bibitem{PhD}
Moghaddam H., PhD Thesis "Rehaeting in Early Universe Cosmology”.
\bibitem{marko}
Marko A., Gasperis G., Paradis G., Cabella P., arXiv:1907.06084.
\bibitem{martin10}
Martin J., Ringeval C., Phys. Rev. D82, 023511, (2010);
Martin J., Ringeval C., Vennin V., Phys. Rev. Lett. 114, no.8, 081303, (2015);
Kawasaki M., Kohri K., Sugiyama N., Phys. Rev. Lett. 82, 4168, (1999).
\bibitem{salas}
Salas P., Lattanzi M., Mangano M., Miele G., Pastor S., Pisanti O., PRD 92, 12, 123534 (2015).
\bibitem{TR}
Kawasaki M., Kohri K., Moroi T., PRD 63, 103502, (2001).
\bibitem{gravitino}
Giudice G., Tkachev I., Riotto A., JHEP 9908, 009, (1999); Buchmuller W., NPB 606, 518, (2001).
\bibitem{TB90}
Traschen J., Brandenberger R., PRD 42, 2491, (1990).
\bibitem{KLS94}
Kofman L., Linde A., Starobinski A., Phys. Rev. Lett. 73, 3195, (1994).
\bibitem{KLS97}
Kofman L., Linde A., Starobinski A., Phys. Rev. D 56, 3258, (1997).
\bibitem{Boy95}
Boyanovsky D., et. al., PRD 52, 6805, (1995); PRD 51, 4419, (1995).
\bibitem{Al11}
Allahverdi R., et. al., Phys. Rev. D83, 123507, (2011).
\bibitem{FKL98}
Felder G., Kofman L., Linde A., Phys. Rev. D59, 123523, (1999).
\bibitem{F95}
Fujisaki H., et. al., Phys. Rev. D53, 6805, (1996).
\bibitem{ABM00}
Allahverdi R., Bastero-Gil M., Mazumdar A., Phys. Rev. D64, 023516, (2001).
\bibitem{AIP}
Kirilova D., Panayotova M., AIP Conf. Proc. 2075, 090017, (2019).
\bibitem{BS}
Kirilova D., Panayotova M., Publ. Astron. Soc. "Rudjer Bockovic", Proc. XII SB
Astronomical Conference, 25-29 September 2020, Sokobanja, Serbia (to be published).
\bibitem{DL82}
Dolgov A., Linde A., Phys. Lett. 116B, 329, (1982).
\bibitem{SV84}
Shafi O., Vilenkin A., Phys. Rev. Lett. 52, 691, (1984).
\bibitem{AFW82}
 Abbott L., Fahri E., Wise M., Phys. Lett. B117, 29, (1982).
\bibitem{Bruck16}
Van de Bruck C., Dansby P., Paduraru L., International Journal of Modern Physics D 26, 13, (2013).
\bibitem{A06}
Allahverdi A., et. al., Phys. Rev. Lett 97, 191304, (2006).
\bibitem{F17}
Ferrantelli A., Eur. Phys. J. C 77, 716, (2017).
\bibitem{NOS83}
Nanopoulos D., Olive K., Srednicki M., Phys. Lett. B127, 30, (1983).



\end{thebibliography}
\end{document}